\documentclass[reprint,aps,prl,twocolumn,superscriptaddress]{revtex4-1}
\usepackage{graphicx}
\usepackage{mathrsfs}
\usepackage{bm}
\usepackage{hyperref}
\usepackage{dcolumn}
\usepackage{amsmath}
\usepackage{amssymb}
\usepackage{bbold}
\usepackage{mathtools}
\usepackage{color}

\newcommand{\br}{\mathbf{r}}

\newcommand{\bS}{\mathbf{S}}
\newcommand{\bB}{\mathbf{B}}

\usepackage[normalem]{ulem}

\begin{document}

%\title{Terahertz Magnetometry with Defect in van der Waals Monolayers}
\title{All-optical magnetic imaging with spin defects\\ in van der Waals materials at Angstrom-scale}
% \title{All-optical Magnetometry with spin defects in van der Waals materials at Angstrom-scale}

\author{Ning Wang}
\email{ningwang@hust.edu.cn}
\author{Jianming Cai}
\email{jianmingcai@hust.edu.cn}
\affiliation{School of Physics, Hubei Key Laboratory of Gravitation and Quantum Physics, Institute for Quantum Science and Engineering, Huazhong University of Science and Technology, Wuhan 430074, China}
\affiliation{International Joint Laboratory on Quantum Sensing and Quantum Metrology, Huazhong University of Science and Technology, Wuhan 430074, China}
% \author{Di Xiao}
% \affiliation{Department of Physics, University of Washington, Seattle, Washington 98105, USA}
% \affiliation{Department of Material Science and Engineering, University of Washington, Seattle, Washington 98105, USA}
% \author{Allan H. MacDonald}
\author{Chao Lei}
\email{leichao.ph@gmail.com}
\affiliation{Department of Physics, The University of Texas at Austin, Austin, Texas 78712, USA}

\begin{abstract}
    Magnetic imaging with ultra-high spatial resolution is crucial to exploring the magnetic textures of emerging quantum materials. We propose a novel magnetic imaging protocol that achieves Angstrom-scale resolution by combining spin defects in van der Waals materials and terahertz scattering scanning near-field optical microscopy (THz s-SNOM). Spin defects in the atomic monolayer enable the probe-to-sample distance diving into Angstrom range where the exchange interactions between the probe and sample spins become predominant. This exchange interaction leads to energy splitting of the probe spin in the order of millielectronvolts, corresponding to THz frequencies. With THz optics and the spin-dependent fluorescence of the probe spin, the interaction energy can be resolved entirely through optical methods. Our proposed all-optical magnetic imaging protocol holds significant promise for investigating magnetic textures in condensed matter physics due to its excellent compatibility and high spatial resolution.
\end{abstract}

\maketitle

\textit{Introduction---} Detecting spin textures is crucial to investigating collective phenomena such as magnetic ordering, nontrivial topological effects, and phase changes in novel condensed matter systems\cite{Huang2017, Gong2017, Qi2011, Hasan2010, Coey2013, Santamaria2013, Dieny2017}. A comprehensive understanding of magnetism at the nanometer scale necessitates resolving detailed information about the spatial arrangement of spins at the atomic scale. Magnetic imaging techniques with angstrom resolution have thus been developed, including spin-polarized scanning tunneling microscopy (STM)\cite{Wiesendanger1990Observation, Wiesendanger2009Spin}, electron spin resonance (ESR) STM\cite{Baumann2015Electron, Willke2018Hyperfine, Yang2019Coherent}, and magnetic exchange force microscopy (MExFM) \cite{Rugar2004Single, MExFM2007}. In these techniques, spin-polarized and ESR STM were designed based on the spin-dependent electronic tunneling current, and the spin-polarized tip in ESR STM not only serves as a probe but also affects the coherence of spins in the sample\cite{Yang2019Coherent}. MExFM depends on the mechanic resonance and magnetic tip may disturb the sample.

In this letter, we propose a novel magnetic imagining protocol by integrating spin defects in monolayer van der Waals (vdW) material with terahertz scattering scanning near-field optical microscopy (THz s-SNOM). We show that when the probe-to-sample distance is down to ten-angstrom scale, the dominant interactions are magnetic exchange interactions with exponential dependence on the distance between the probe and sample spins. At this region, the energy-splitting of spin defects is order of milivolts, corresponding to THz resonance frequencies, which can be detected by combining the THz optics and the spin-dependent fluorescence of spin defects. Based on this protocol, we estimate the achieved spatial resolution could be improved to the Angstrom scale.

Spin defects in two-dimensional vdw materials such as $\mathrm{MoS_{2}}$\cite{Ye2019Spin, Aharonovich2016Solid}, $\mathrm{WSe_{2}}$\cite{Aharonovich2016Solid, Chakraborty2015Voltage} and hBN\cite{gottscholl2020,gottscholl2021,stern2022,haykal2022,ramsay2023} have been extensively explored recently, especially in the monolayer. Compared with spin defects in 3D structures, e.g. nitrogen-vacancy (NV) centers in diamond, defects in 2D materials allow probe-to-sample distance down to the angstrom scale and have the flexibility for integration with low-dimensional devices. Among these, much progress has been made in manipulating spin states of defects in hBN, especially for the negatively charged boron vacancy ($V_\mathrm{B}^{-}$) centers \cite{gottscholl2020,gottscholl2021,stern2022,haykal2022,ramsay2023}, which have been explored for sensing magnetic fields, spin fluctuations, strain, and temperature etc\cite{vaidya2023,gottscholl2021spin,huang2022,lyu2022,robertson2023,rizzato2023,sasaki2023}. 
Here, we employ $V_\mathrm{B}^{-}$ centers as a spin defect probe for our protocol. 

\begin{figure}[ht!]
  \centering
  \includegraphics[width=1.0 \linewidth ]{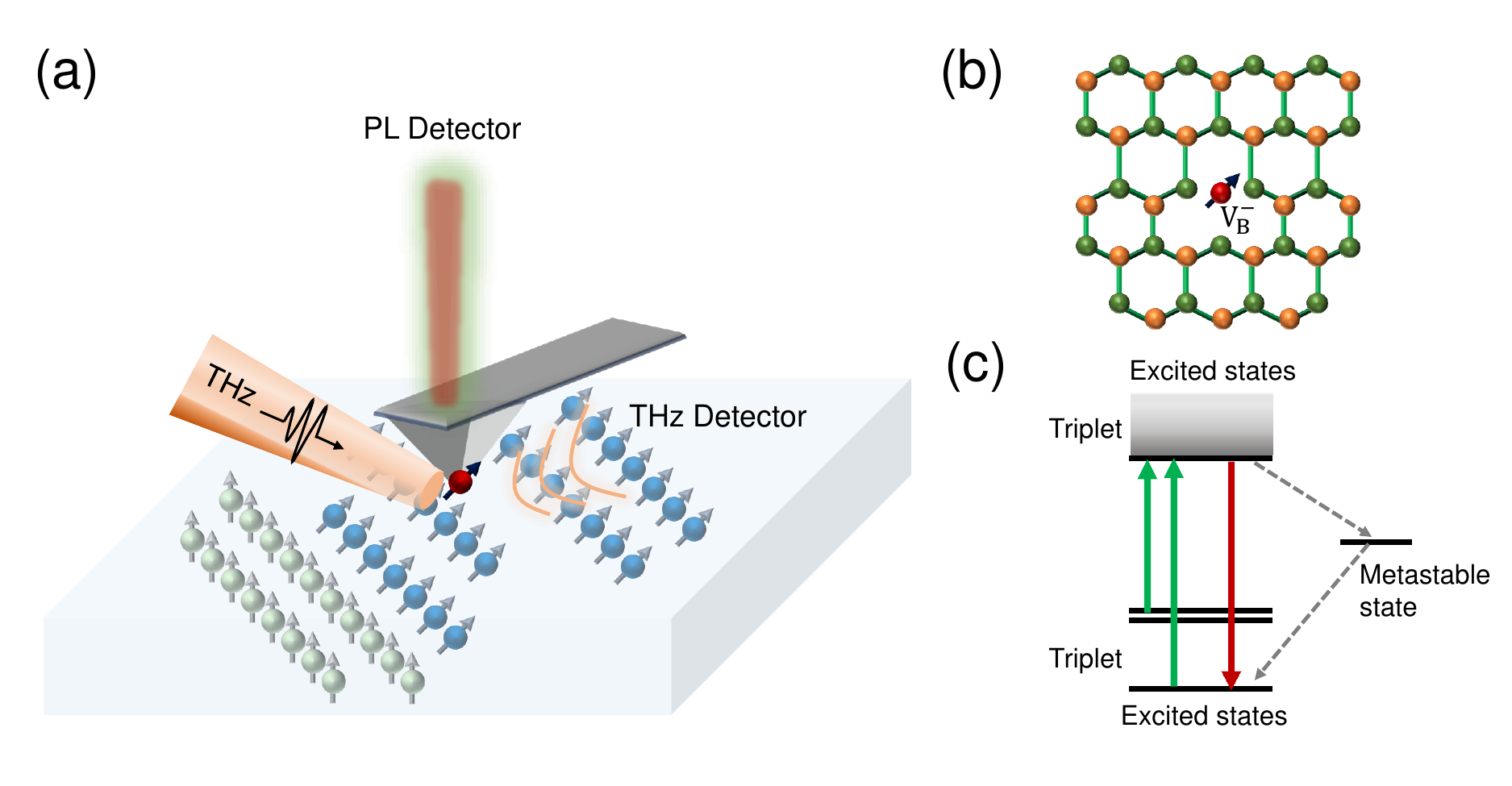}
  \caption{Illustration of single defects spin magnetometer coupled to THz s-SNOM.
  (a) Scheme of proposed magnetic imaging protocol. THz s-SNOM is combined with single $V_\mathrm{B}^{-}$ in monolayer hBN (located at the scanning tip apex).
  (b) Crystal structure of hBN and the corresponding $V_\mathrm{B}^{-}$ center, which is a negatively charged vacancy of boron atoms.
  (c) Electronic structure of $V_\mathrm{B}^{-}$ center.
  }\label{fig:scheme}
\end{figure}

\begin{figure}[h!]
  \centering
  \includegraphics[width=0.9 \linewidth ]{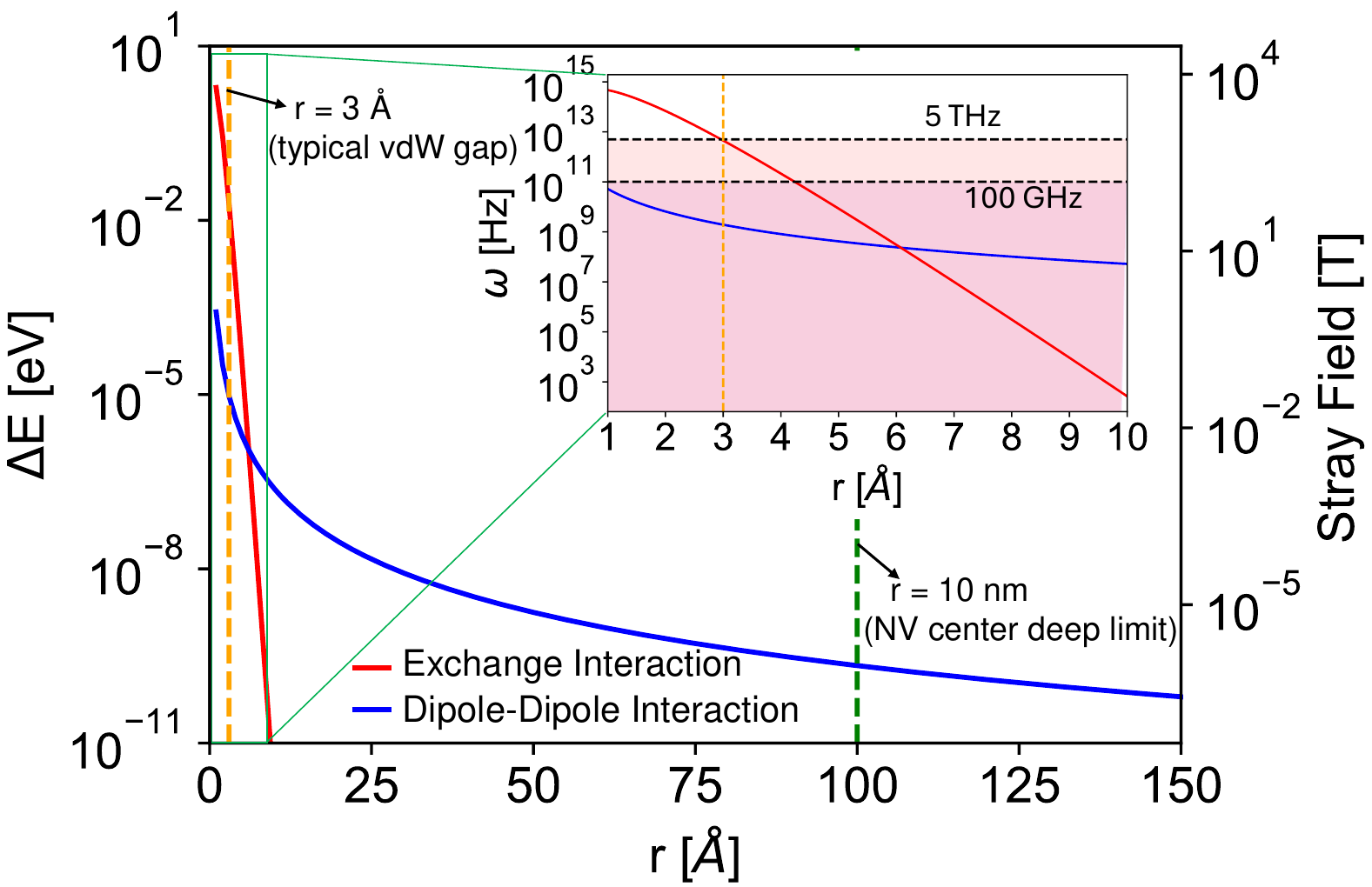}
  \caption{Magnetic interactions versus probe-to-sample distance. The red curves correspond to the energy splitting (labeled on the left axis) due to the magnetic exchange interactions, and the blue curves are the plots of magnetic dipole-dipole interaction. The magnetic stray field at the tip from the sample spin is labeled on the right axis, where we set the direction between the tip and sample spins along the spin directions. In the insert, we plot the corresponding resonance frequency versus the distance. The typical van der Waals gap is labeled with a yellow dashed line, whereas the depth limit of the NV center from the diamond surface is labeled with a green vertical line. 
  }\label{fig:magnetic}
\end{figure}

\textit{Single spin magnetic interactions---}
The detection of spins mostly depends on magnetic dipole-dipole interactions, spin-dependent tunneling of electrons, direct magnetic exchange interactions, etc. Single spin defects in monolayer vdW materials probe sample spins mainly by two kinds of magnetic interactions: magnetic dipole-dipole interactions and magnetic exchange interactions. Although the hBN flakes with $V_\mathrm{B}^{-}$ centers utilized were typically several tens of nanometers thick, which is far away from the 2D limit, recent experimental work by Durand et al.\cite{Durand2023} demonstrated that $V_\mathrm{B}^{-}$ centers in the few-atomic-layer hBN flakes function well and thus shows great progresses toward single $V_\mathrm{B}^{-}$ centers in monolayer hBN. In the following, we assume that the $V_\mathrm{B}^{-}$ center is located at monolayer hBN to illustrate our scheme. To investigate the magnetic interactions between probe and sample spins versus the distance, a spin-1 \cite{gottscholl2020} Hamiltonian is employed to model the ground states of $V_\mathrm{B}^{-}$ color center under an external field, which reads as:

\begin{equation}\label{Ham_VBminus}
     H_{\text{$V_\mathrm{B}^{-}$}} = D (S_z^2 -S(S+1)/3) + g \mu_{\text{B}} \bB_{ext} \cdot \bS.
\end{equation}
Here $D\simeq14.4\mathrm{\mu eV}$ is the zero-field splitting parameter (strain effects can be ignored approaching monolayer limit\cite{Durand2023}), $\bS$ is the electron spin operator, $g$ is the g-factor, $\mu_{\text{B}}$ is the Bohr magneton, and $\bB_{ext}$ is the external magnetic field generated at the spin position. For magnetic dipole-dipole interactions between $V_\mathrm{B}^{-}$ centers and spins in the sample, which are long-range interactions with the Hamiltonian as:
\begin{equation}
    H_{DD} = \frac{-\mu_0 \gamma_1 \gamma_2 \hbar^2}{4 \pi |\br|^3} (3(\bS_1 \cdot \hat{\br})(\bS_2 \cdot \hat{\br})-\bS_1 \cdot \bS_2), 
\end{equation}
where $\mu_0$ is the vacuum magnetic permeability, $\gamma_i \equiv g_i \mu_{\text{B}}/\hbar$ is the gyromagnetic ratios of two particles with spin quanta $S_1$ and $S_2$ and $\hbar$ the reduced Plank constant, $\hat{\br}$ is a unit vector parallel to the line joining the centers of the two dipoles $\bS_1$ and $\bS_2$, $|\br|$ is the distance between the two dipoles. In the context of magnetic dipole-dipole interactions, the spins within the sample can be probed and quantified through the detection of the stray field:
\begin{equation}
    \bB(\br) = \frac{-\mu_0 \gamma \hbar}{4 \pi |\br|^3} (3 \hat{\br} (\bS \cdot \hat{\br})- \bS).
\end{equation}
The magnetic dipole-dipole energy and corresponding stray magnetic field versus the distance $r$ between spins are plotted with the blue curve in Fig. \ref{fig:magnetic}. Since the stray field can be tens of $\mathrm{\mu T}$ and up to several $\mathrm{mT}$ when $r \to 10~\mathrm{nm}$, it can thus be easily detected by $V_\mathrm{B}^{-}$ centers through optical detected magnetic resonance measurements (ODMR) by sweeping the microwave frequency. 

When the probe-to-sample distance decreases to the angstrom scale, the magnetic exchange interactions can not be omitted and act as:
\begin{equation}
    H_{ex} = J_{ex} \bS_1 \cdot \bS_2,
\end{equation}
with $J_{ex}$ the exchange interaction constant that depends exponentially on the spin-spin distance \cite{Herring1964}:
\begin{equation}
    J_{ex}(r) \approx 1.641 \frac{e^2}{2a_{_B}} (\frac{r}{a_{_B}})^{5/2} e^{-2r/a_{_B}}.
\end{equation}
Here $a_{_B}$ is the Bohr radius. The dependence of energy splitting contributed by the magnetic exchange interaction versus the distance $r$ is plotted with the red curve in Fig. \ref{fig:magnetic}, indicating a dominant role when the distance decreases to several angstroms. 

As shown in the insert of Fig. \ref{fig:magnetic}, the energy splittings have resonant frequencies up to 5 THz when the spin distance approaches around 3 $\AA$. Due to the exponential dependence of exchange interaction on distance, the resonant frequency decreases to around 20 GHz when the probe-to-sample distance increases to around 5 $\AA$. The corresponding exchange splitting energies thus range from around  0.02 meV to 20 meV, this order is consistent with the density functional theory calculations \cite{Shen2024}. Since the resonance frequencies are in the range of THz, they can be resolved by leveraging state-of-the-art terahertz techniques and the unique properties of spin defects in monolayer 2D vdW materials, which will be discussed in the following section.

\textit{Magnetic imaging via integration of spin defects with THz s-SNOM---}
To fully demonstrate the excellent spatial resolution of the defect center in monolayer 2D vdW materials, the defect center can be integrated with the THz s-SNOM illustrated in Fig.1\label{fig:scheme} (a). THz s-SNOM is a powerful THz imaging technique recently developed to enhance the spatial resolution of terahertz spectroscopy \cite{Cocker2021Nanoscale}. In traditional far-field THz studies, spatial resolution is constrained by diffraction, which limits its applicability in nanoscience. However, THz s-SNOM addresses this fundamental limitation by utilizing the strong confinement of optical fields at the apex of a sharp metal tip. As a result, the spatial resolution is determined by the size of the tip rather than the wavelength employed. The strength of the scattered THz field at the interface provides insights into the refractive index and hence the related conductivity and properties of charge carriers of the sample, thus it is widely used in high-resolution THz imaging and conductivity investigations \cite{Cocker2021Nanoscale,zhang2018terahertz,maissen2019probes, THzsnorm2024}.

In our protocol, spin defects are integrated into the THz s-SNOM. This requires a specialized atomic force microscopy (AFM) tip, as illustrated in Fig. 3 (a)\label{fig:TDTHZ}. A single defect center, such as a single $V_\mathrm{B}^{-}$ center, is supposed to be located at the apex of the AFM tip.  Diamond tip with single NV centers at the apex of the tip has already been successfully fabricated in NV scanning magnetometry \cite{Appel2016Fabrication,Zhou2017Scanning,Siyushev2019Photoelectrical,Huxter2023Imaging,Song2021Direct}. But fabrication of hBN tip is quite challenging because of the difficulties associated with growing large-sized hBN crystals. This issue can be addressed by either carefully attaching an hBN nanoflake containing $V_\mathrm{B}^{-}$ to the tip apex\cite{Tetienne2014Nanoscale} or by growing a thin layer of hBN on the tip apex, ensuring that a single $V_\mathrm{B}^{-}$ is present on the surface layer. Furthermore, to initialize and read out the spin states of $V_\mathrm{B}^{-}$ centers, it is essential for both the cantilever and tip to be transparent to the fluorescence emitted by the $V_\mathrm{B}^{-}$ centers as well as the excitation green laser. AFM tips made of diamond or silicon oxide meet this requirement \cite{Appel2016Fabrication,Zhou2017Scanning,Siyushev2019Photoelectrical,Huxter2023Imaging,Song2021Direct}.
Since an insulator-based tip exhibits limited enhancement of the THz field at its apex (see details in supplementary material), a 5 nm layer of gold (Au) can be coated around the surface to concentrate the THz field, thus enhancing the coupling between the THz radiation and the spin defects. The THz field enhancement at the tip's apex is about 40 times according to our simulation with 5 nm coated Au, as shown in Fig.3 (b) \label{fig:TDTHZ}, assuming the incident angle between the incident THz light and the horizontal plane is 45 degrees.

During the scanning, a constant-distance scheme can be employed for magnetic imaging, i.e. the separation between the tip and the sample is maintained at a fixed distance. This configuration can be achieved in either contact mode or tapping mode, both of which are standard operational modes in conventional AFM. A critical component of this scheme is a frequency-tunable THz source. Significant advancements have been made to generate THz radiation with arbitrary high frequency with substantial energy output, including THz comb\cite{Lu2019Room}, photomixing\cite{Safian2019Review}, and so on. To precisely measure the exchange energy, the frequency step of the THz radiation changes should be in the order of tens of MHz for $V_\mathrm{B}^{-}$ centers, which can be realized by combining photomixing of two lasers. A green laser is used to polarize and readout the spin states, and the photoluminescence emitted by the spin defects can be recorded simultaneously when the frequencies of the THz light source are varied. This allows us to obtain the resonant interaction energy, similar to ODMR measurements conducted in the microwave range. Consequently, during the measurement of the interaction energies, all techniques are optical and the detection system is significantly simplified.

\begin{figure}[h!]
  \centering
  \includegraphics[width=0.9 \linewidth ]{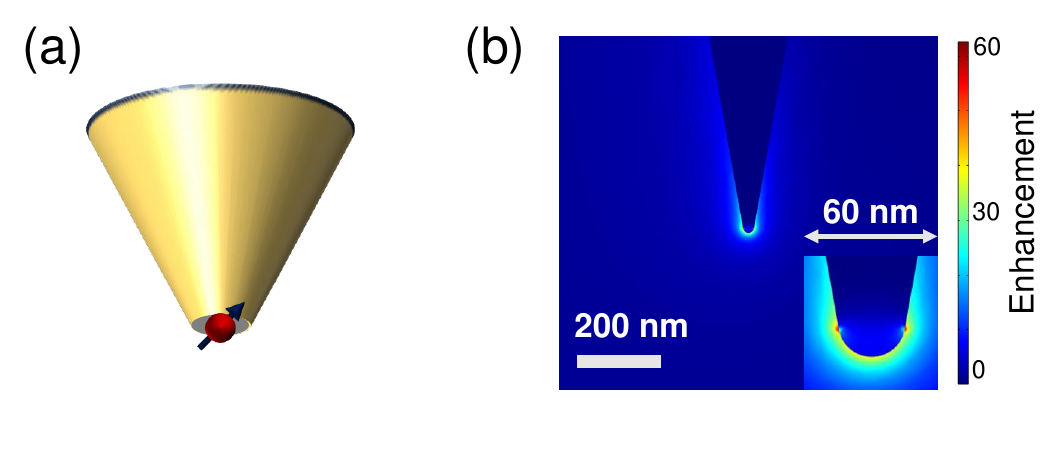}
  \caption{Enhancement of the THz field by tip. (a) The structure of the scanning tip. The tip is made of silicon oxide or diamond with 5 nm Au coated around the surface and a single $V_\mathrm{B}^{-}$ center at the apex of the tip (b) The distribution of THz field enhancement at the apex of the tip, assuming the apex radius is 15 nm and the incident angle of the THz light is 45 degrees.
  }\label{fig:TDTHz}
\end{figure}

\begin{figure}[h!]
  \centering
  \includegraphics[width=0.9 \linewidth ]{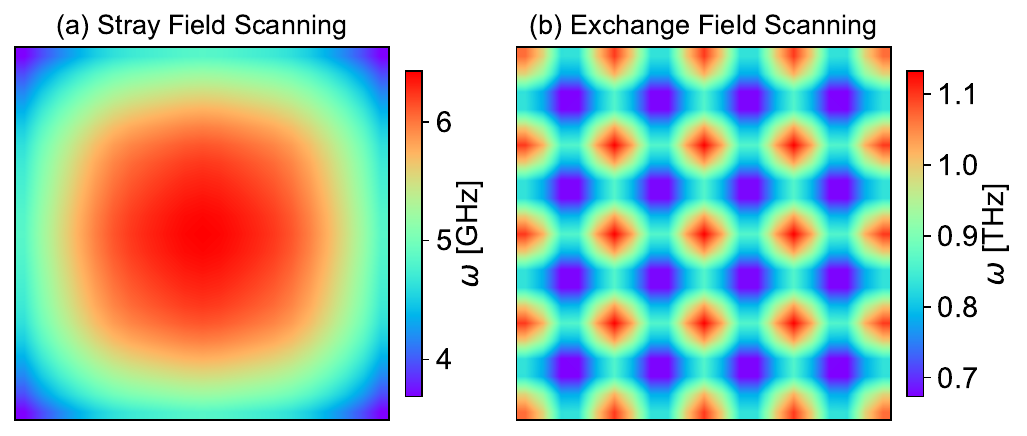}
  \caption{Simulated scanning resonance frequency of the tip spin when it is scanned through a $5 \times 5$ square lattice, with a lattice constant of 3 $\AA$, and the vertical distance between the tip and sample is set to be 4 $\AA$. 
  (a) is for the stray field scanning and 
  (b) is for the magnetic exchange field scanning.
  }\label{fig:scanning}
\end{figure}

\textit{Magnetic imaging of spin textures---}
In this section we will demonstrate the performance of our proposed magnetic imaging protocol and compare it with NV stray-field scanning magnetometry based on dipole-dipole interactions. In NV stray-field scanning, the total Zeeman energy at the tip is contributed from all the spins in magnetic systems, whose Hamiltonian is
\begin{equation}
    H_{DD}^{tip} = \sum_{i} \frac{-\mu_0 \gamma_t \gamma_i \hbar^2}{4 \pi |\br_i|^3} (3(\bS_t \cdot \hat{\br_i})(\bS_i \cdot \hat{\br_i})-\bS_t \cdot \bS_i), 
\end{equation}
where $\bS_t$ labels the spin in the tip, and $\br_i$ is the spatial vector that connects the spin in the sample and tip. The Zeeman shift with a probe-sample spin distance of around 10 nm is in order of 10 MHz (with FWHM around 1 MHz)\cite{casola2018, Ngambou2022}. However, since magnetic dipole-dipole interaction is a long-range interaction, it is more suitable for detecting a single spin signal. In condensed matter systems, the magnetic structures are mostly atomic scale, the signal from the magnetic dipole-dipole field usually needs the reconstruction process to determine the microscopic spin structures, which is not unique. The reconstruction is not able to improve the spatial resolution even if the tip moves close to the sample, as shown in Fig. \ref{fig:scanning} (a), in which we plot the energy splitting (resonance frequency) of the tip spin contributed by magnetic dipole-dipole interactions when it scans a $5 \times 5$ square lattice with a lattice constant of 3 $\AA$, the vertical distance between the tip and sample is set to be 4 $\AA$.

When consider the magnetic exchange interaction, the Hamiltonian of the tip reads:
\begin{equation}
    H_{ex}^{tip} = \sum_{i} J_{ex}^{i} \bS_t \cdot \bS_i.
\end{equation}
Here $J_{ex}^{i}$ is the magnetic exchange coupling between the i$^{th}$ spin of the sample and tip. With the constant-distance scheme metioned above, the interaction energy could be directly measured.
As shown in Fig. \ref{fig:scanning}(b). the measured resonance frequencies of the tip spin vary from 0.7 THz to 1.1 THz
when we set the same lattice constant and tip-sample spin distance as the stray field scanning in Fig. \ref{fig:scanning}(a).
In the simulations, the defects in hBN monolayer are assumed to be put close to the sample surface; a vertical distance of 4 $\AA$ is in order but a bit larger than the typical vdW gap. With this constant-distance scanning, the spatial resolution can reach to atomic scale.

In addition, the magnetic imaging mode can be modified to the resonant-frequency mode, which requires only a single THz source with a fixed frequency. When the THz field is resonant with the energy splitting of the spin defects, the photoluminescence decreases significantly. We can maintain this reduced photoluminescence by varying the distance between the sample and the tip\cite{Tetienne2014Nanoscale}; however, the reconstruction of the spin textures is more complex compared to the constant-distance mode. 

\textit{Discussion---}
The proposed angstrom-scale magnetic imaging approach harnesses the unique properties of spin defects and THz technology, which is also fully compatible with scattering THz field detection in THz s-SNOM, enabling THz imaging and THz time-domain spectroscopy. The semiconductive and non-magnetic tip minimizes the perturbation to the sample, ensuring the study of the intrinsic magnetism. Unlike STM and MExFM, the signal contrast in our approach is solely determined by the fluorescence contrast under THz manipulation of the spin defects, independent of the tip-sample distance or the spin interactions, resulting in a significantly more robust signal. Additionally, the all-optical manipulation and detection, along with the absence of requirements for large external magnetic fields, cryogenics, and complicated microwave or electronic setups, greatly reduces the system's complexity.

Nonetheless, our approach remains challenging. As with all magnetic imaging techniques capable of achieving atomic resolution, maintaining low variations and ensuring flat sample surfaces are essential prerequisites. Substantial efforts are still required to achieve the ultimate goal of identifying individual spin defects in monolayers. We believe the integration of THz s-SNOM and spin defects in van der Waals monolayers will serve as a versatile tool for investigating condensed matter physics, enabling the exploration of intrinsic magnetic properties in novel correlated conductive, insulating, and semiconductive magnetic systems.

\textit{Acknowledgements.}
This work was supported by the Innovation Program for Quantum Science and Technology 2023ZD0300600, National Natural Science Foundation
of China 12304571.

%\bibliography{vdw}

%

\end{document}